\documentclass[a4paper,12pt]{article}
\usepackage[cyr]{aeguill}
\usepackage[latin1]{inputenc}
\usepackage[english]{babel}

\usepackage{a4wide}
\usepackage{color}
\usepackage{courier}
\usepackage{url}
\usepackage[pdftex]{graphicx}
\usepackage{caption}
\usepackage{listings}
\usepackage{appendix}
\usepackage[bookmarks=true]{hyperref} 

\newtheorem{wfr}{Rule}

\newcommand{\bs}{\vspace{3mm}} 

\title{Well-formedness and typing rules for UML Composite Structures}
\author{Iulia Dragomir and Iulian Ober\\
\small IRIT - University of Toulouse\\
\small 118 Route de Narbonne, 31062 Toulouse, France\\
\small \{iulia.dragomir, iulian.ober\}@irit.fr 
}
\date{}
 
\begin{document}

\maketitle
\begin{abstract}
Starting from version 2.0, UML introduced hierarchical composite structures, which are an expressive way of defining complex software architectures, but which have a very loosely defined semantics in the standard. In this paper we propose a set of consistency rules that disambiguate the meaning of UML composite structures. Our primary goal was to have an operational model of composite structures for the OMEGA UML profile, an executable profile dedicated to the formal specification and validation of real-time systems, developed in a past project to which we contributed. However, the rules and principles stated here are applicable to other hierarchical component models based on the same concepts, such as SysML. The presented ruleset is supported by an OCL formalization which is described in this report. This formalization was applied on different complex models for the evaluation and validation of the proposed principles.
\end{abstract}

\section{Introduction}
\label{ch:Intro}

This technical report\footnote{This report represents an excerpt of Iulia Dragomir's Master Thesis defended June 2010 at Universit\'{e} Paul Sabatier Toulouse III, France.} introduces a new version of the OMEGA UML Profile and its formalization, an extension based on composite structures. Composite structures were introduced starting with the version 2.0 in the UML standard \cite{OMGUML2} and represent a big evolution in the representation of complex hierarchical systems. Because the UML standard is under-specified in order to preserve the generality of the language, various ambiguities are introduced in the model when using composite structures. Our purpose is to define an expressive set of notions and principles to clarify the composite structures at the modelling level and also at the execution model level. Our rule set can be applied to other component based systems, like SysML \cite{SYSML11} or MARTE \cite{marte}. All the principles were formalized in OCL in order to catch the most frequent modelling issues regarding composite structures. 

Our interest in composite structures is given by the powerful expressiveness of these constructs when modelling the architecture of hierarchical systems. Common applications are real-time embedded systems which can be found in a large number of domains like avionics, aeronautics, consumer electronics and many others. An important research topic is to prove their safety.

The context of our work is the previous OMEGA UML Profile, dedicated to the specification and validation of real-time embedded systems. This profile is based on a subset of UML 1.4 elements for modelling the structure and the behaviour of a system and has as extensions time modelling and observers (elements that express safety properties of a model). The profile is integrated in a platform, the IFx Toolset, which proposes validation techniques like model-checking, simulation and static analysis via a translation to the intermediate IF representation.

\paragraph{Related work.} The idea of UML composite structures is rooted in previously existing languages, notably ROOM~\cite{Selic*ROOM} and SDL~\cite{SDL00}. However, UML adds much complexity with respect to previous models, e.g., by allowing explicit port behaviour specifications, multiple interfaces per port, typing of connectors with associations, etc. Many problems identified in this paper stem from the added complexity. Potential problems and ambiguities in UML composite structures have previously been discussed by other authors \cite{OliverLuukkala06,CuccuruGR08}.  In \cite{CuccuruGR08}, Cuccuru et al. proposed a set of additional rules meant to further clarify the semantics of UML composite structures. While we fully subscribe to the solutions they propose, some issues remain unsolved, and the present paper is complementary to their solutions.

\paragraph{Structure of the report.} Section~\ref{ch:OM&IFx} presents an overview of the previous version of the OMEGA Profile. Section~\ref{ch:CompStruct} introduces the Composite Structures and the well-formedness and typing rules for these structures are presented in Section~\ref{ch:ExtOM}. Section~\ref{ch:CompStruct2IF} presents the principles for the model transformation to an IF model. Section~\ref{ch:OCLF} contains the OCL formalization for profile's rule set and Section~\ref{ch:eval} describes the evaluation of the formalization on a complex model before concluding.

\section{Overview of the OMEGA UML Profile}
\label{ch:OM&IFx}

The OMEGA Profile (\cite{OberGO062}) identifies a subset of the UML language which is sufficiently expressive for modeling the structure and behavior of real-time systems, and for which an operational semantics is defined by closing the relevant semantic variation points left open in the UML standard (\cite{DBLP:journals/scp/DammJPV05}). This profile it is integrated within a framework (the IFx toolset \cite{BozgaGOOS04}) that supports techniques like static analysis, model checking and simulation for validating real-time embedded system. 

The previous version of the OMEGA UML Profile is based on a subset of UML 1.4. From structural point of view the profile consists in :
\begin{itemize}
\item \textbf{Classes.} They can be \textit{active} or \textit{passive}, partitioning the object space in \textit{activity groups}. Each instance of an \textit{active} class\footnote{Active classes are represented with a thick border to distinguish them for passive classes.} defines an \textit{activity group}. Each instance of a \textit{passive} class belongs to one single \textit{activity group}, the one that has created it. They can own attributes, relationships, operations and state machines. Activity groups are considered concurrent and they react to external stimuli (like signals and operation calls) in a run-to-completion manner. When a request is received from the outside environment, it is stored in group's queue and is handled later when the group is stable. By \textit{stable} we mean that every object owned by the group has no spontaneous transitions (transitions that are guarded by a boolean condition and have no trigger) or pending operations from inside the group (i.e., the \textit{object} is \textit{stable}).
\item \textbf{Structural features.} Classes have attributes which can have \textit{predefined types}: Integer, Real, Boolean, or \textit{reference types}. Since the OMEGA UML Profile was developed for modelling real-time embedded systems, two extensions representing time (\textit{timer} and \textit{clocks}) have been included in the profile. \textit{Timer} objects measure durations. They may be set to a relative deadline and can be reset. Upon deadline different operations may be executed by objects: sending a signal, calling an operation, etc. \textit{Clock} objects measure also durations, but their values can be consulted by other objects. 
\item \textbf{Relationships.} The relationships that can be defined between classes are \textit{associations} and \textit{generalisations}. The associations supported are \textit{simple} or \textit{compositional}. 
\end{itemize}

From behavioural point of view, the OMEGA UML contains:
\begin{itemize}
\item \textbf{Operations.} We distinguish two types of operations: \textit{primitive} and \textit{triggered}. \textit{Triggered operations} are a special kind of transition trigger: the call of such an operation enables the transition which this guards. \textit{Primitive operations} are similar to the methods in object oriented programming languages: they are subject to polymorphism and dynamic binding because of the inheritance relationship that may be defined between classes. They can own a body which is described by an action. When an operation is called by an object from the same activity group, the call is handled immediately by the object called using a call stack. If the call is made from another activity group, then it is queued by the receiving group and handled in a later run-to-completion step.
\item \textbf{Signals.} They are the second method for asynchronous communication between objects and are usually used for triggering actions in the state machine of the target object. They can have parameters and they differ from triggered operation in the sense that they cannot have a return value. Signals always pass through object's activity group and are handled in a later run-to-completion step, no matter if the target is in the same activity group as the sender or not.
\item \textbf{State machines.} They describe the behaviour of a class in term of states, transitions, triggers, actions, etc.
\item \textbf{Actions.} They describe the effect of a transition in a state machine or the body of an operation. The OMEGA UML Profile introduces a textual action language, OMAL, compatible with the action metamodel of UML and which covers notions like: object creation/destruction, operation calls, expression evaluation, variable assignments, signal output, return action and control flow structuring statements (if-then-else and do-while). 
\end{itemize} 

Besides the timing extension, OMEGA UML Profile introduces the notion of \textit{UML Observers}. They are special objects that monitor the system, respectively its states and its events. Observers are modelled by classes stereotyped with \verb+<<observer>>+. They have local memory and a state machine describes their behaviour. The states qualified as \verb+<<error>>+ states can be used in the model in order to express the satisfaction or the non satisfaction of a safety property. Observers may access any part of UML model's state (object attributes and states, signal queues) and they may use clocks to express timing properties. So, special events have been defined for observers in order to meet their purpose, events related to: 
\begin{itemize}
\item signal exchange: \textbf{send}, \textbf{acceptsignal}, \textbf{receivesignal};
\item operation calls: \textbf{invoke}, \textbf{receive} (reception of call), \textbf{accept} (start of the actual processing of call), \textbf{invokereturn} (sending of a return value), \textbf{receivereturn} (reception of the return value), \textbf{acceptreturn} (consumption of the return value);
\item execution of actions or transitions: \textbf{start}, \textbf{end}, \textbf{startend};
\item timers: \textbf{occur}, \textbf{timeout}, \textbf{set}, \textbf{reset}.
\end{itemize} 
The trigger of an observer transition is a \textbf{match} clause specifying the type of the event (previously presented), some related information (for example the operation name) and observer variables that may receive related information (variables receiving the values of the signal/operation call parameters). 

For further details on the time extension and observers the reader is referred to \cite{OberGO062}.

\section{Overview of the Composite Structures}
\label{ch:CompStruct}

Composite structures have been introduced in the UML standard starting with version 2.0. They refer to ``a composition of interconnected elements, representing run-time instances collaborating via communication links to some common objectives'' (\cite{OMGUML2} pp. 161). Composite structures are a big evolution in modelling a system and are often used for the hierarchical  representation of real-time embedded systems.

\begin{figure}[!ht]
\begin{center}
\includegraphics[height=8cm]{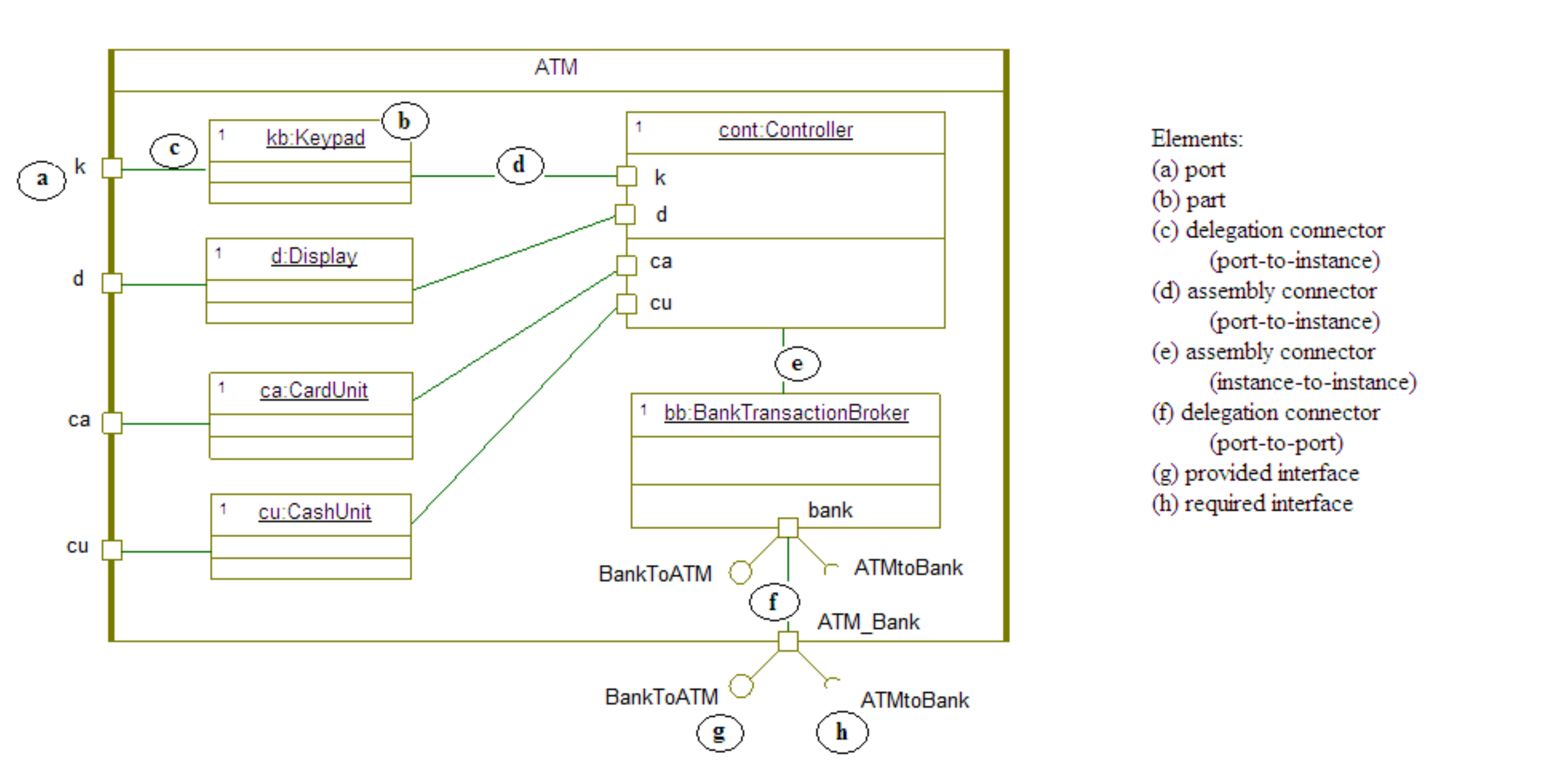}
\caption{Composite structure example}
\label{fig:CSEx}
\end{center}
\end{figure}

A \textit{composite structure} is formed by inner components, that are called \textit{parts}, and communication paths, that are called \textit{connectors} or \textit{links}. \textit{Parts} are instances of classes with predefined role and they usually are in a fix number within the composite structure. In Figure~\ref{fig:CSEx} parts are represented by the instances of Keypad, Display, CashUnit, CardUnit, Controller and BankTransactionBroker. \textit{Links} connect inner components (e.g. \textbf{e} in Figure~\ref{fig:CSEx}), an inner component with a port (e.g. \textbf{c, d}) or two ports (e.g. \textbf{f}), so that these elements can communicate between them via signals or operation calls. Such links can transport signals between elements that know how to answer to them. A link can be the realization of an \textit{association} (especially in the case of a connector between two parts), but this is not mandatory. The UML standard classifies links in two categories: \textit{delegation links} which connect the composite structure with one of its components (part or port of a part) and \textit{assembly links} which connect two components between them. \textit{Delegation links} can be separated in \textit{outbound delegation links} and \textit{inbound delegation links}, depending of connector's direction (if it is oriented to the outside environment or, correspondingly, to the inner structure).

A \textit{port} (e.g. the elements \textbf{k, d, ca, cu} from ATM and Controller, \textbf{bank} from BankTransactionBroker and \textbf{ATM\_Bank} from ATM in Figure~\ref{fig:CSEx}) is an interaction point between its owner and the outside environment. Every port has a contract, given by classifiers (interfaces or classes), which allows it to handle known requests by forwarding them in the needed direction. These requests can be incoming requests from the environment (the port \textit{provides} the interface; e.g. \textbf{g} from Figure~\ref{fig:CSEx}) or outgoing requests to the environment (the port \textit{requires} the interface; e.g. \textbf{h} from the same Figure).\footnote{\textit{Provided} and \textit{required} are defined from the component owning the port point of view.} 

Anyhow composite structures, as presented in UML standard, are ambiguous. For example, every connector links two entities which can be either ports or components (parts). In both cases, the two entities are typed\footnote{For a part the type is given by the class whose instance it is and for a port the type is given by its contract.}. In addition, the modeller can specify that the connector \textit{realizes} an \textit{association}. It is clear that, in general, connecting entities of arbitrary types does not make sense, and there should be clear compatibility rules (based on types, link direction, etc.) specifying what are the well formed structures. However, these type compatibility rules for connectors are not detailed in UML. The standard merely states that ``the connectable elements attached to the ends of a connector must be compatible.'' and that ``what makes connectable elements compatible is a semantic variation point'' (\cite{OMGUML2} pp. 175-176). Various causes of ambiguity, such as the existence of several connectors starting from a same end-point, are not even mentioned. 

Our purpose is to define a rule set in order to disambiguate the composite structures so that we shall have a clear a coherent executable semantics. Therefore, we had extended the OMEGA UML Profile to cover unambiguous composite structures by setting well-formedness constraints and by clarifying the run-time behaviour of these structures. A formalization of the rules provided is needed for proving the type-safety of our system.

\section{The extended OMEGA Profile: well-formedness and typing rules}
\label{ch:ExtOM}

The extended OMEGA UML Profile (called OMEGA2) introduces an expressive and unambiguous set of constructs for modelling hierarchical structures, with an operational semantics that integrates in the existing execution model of OMEGA. The typing system and the consistency rules we have formulated can be applied to other component-based models like SysML (\cite{SYSML11}) or MARTE (\cite{marte}).

\subsection{Bidirectional ports}
\label{sec:BidirPort}

\begin{figure}[!ht]
\begin{center}
\includegraphics[width=7cm]{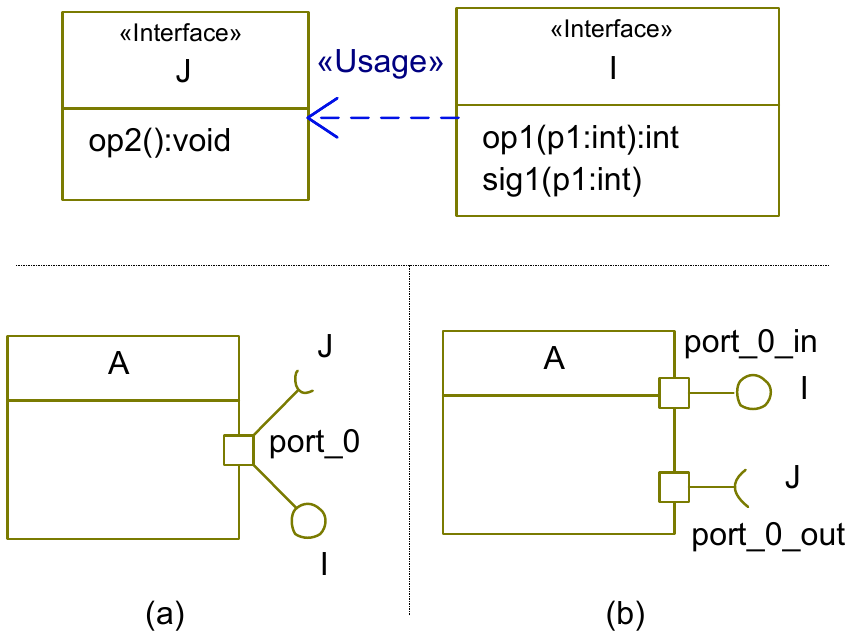}
\caption{(a) - Example of a biriderctional port, (b) - Equivalent in OMEGA2}
\label{fig:bidirPort}
\end{center}
\end{figure}

A first typing problem comes from the fact that in UML the ports are bidirectional, i.e. they can specify a set of allowed incoming requests (the \textit{provided} interfaces) and a set of allowed  outgoing requests (the \textit{required} interfaces). This is represented in the model as follows: all the interfaces that are directly or indirectly \textit{realized} by the type of the port (its contract) are considered to be \textit{provided} interfaces. The \textit{required} interfaces are those interfaces for which there exists a \textit{Dependency} stereotyped with \verb+<<Usage>>+ between the port type (or one of its supertypes) and the respective interface(s). Figure~\ref{fig:bidirPort}-a shows a simple example of bidirectional port.

The type of the port \verb+port_0+ in Figure~\ref{fig:bidirPort}-a is $I$. However, the fact that the port is bidirectional raises typing problems, which are apparent in the following situations:
\begin{itemize}
\item When \verb+port_0+ is used by $A$ to send out requests conforming to interface $J$, by an action such as ``\verb+port_0.op2()+''. In this case, \verb+port_0+ has to be treated by the type system as an entity of type $J$, although it is declared of type $I$.
\item When one wants to specify behaviour of \verb+port_0+ by a state machine\footnote{This is deemed possible by the UML 2.x standard \cite{OMGUML2}, but without further detail.}. Then the state machine has to handle requests coming from both directions, i.e. requests conforming both to $I$ and to $J$.
\end{itemize}

These typing inconsistencies are not addressed by the UML standard. When we translate UML models into their IF description (or other implementation languages) they raise homologous problems for the typing of the actual object that will represent the port. A general solution, based on qualifying the types ($I$, $J$) with the corresponding directions ($in$, $out$) and on allowing the port entity to comply to multiple types, is possible but it greatly complicates the type checking of UML models.

For these reasons, the solution we adopt in OMEGA2 is to forbid bidirectional ports. This is possible because any bidirectional port can be split in two unidirectional ports, like in the example from Figure~\ref{fig:bidirPort}-b, although it can be argued that it leads to less convenient models.

Syntactically, an unidirectional outgoing port specifying a \textit{required} interface $J$ (such as \verb+port_0_out+ from Figure~\ref{fig:bidirPort}-b) will be represented as a port typed with $J$ and stereotyped with \verb+<<reversed>>+ (to distinguished it from a port \textit{providing} $J$).\footnote{Note that a mechanism identical to the \texttt{<<reversed>>} stereotype is supported by the IBM Rhapsody tool~\cite{Rhapsody}, including support for graphical representation using the standard \textit{required interface} symbol of UML like in Figure~\ref{fig:bidirPort}-b. For editing convenience, the Rhapsody representation is also supported by the IFx2 tools.}

\subsection{Directionality rules}
\label{sec:DirR}

A second typing problem is raised by connectors. No compatibility rules for links are given by the standard. Before presenting type compatibility issues for links, some simple directionality rules must be observed by well-formed structures:

\begin{wfr}
\label{wfr:dir1}
If a \emph{delegation} link exists between two ports, the  direction \emph{(provided} or \emph{required)} of the ports must be the same.
\end{wfr}

\begin{wfr}
\label{wfr:dir2}
If an \emph{assembly} link exists between two ports, one of the ports (the source) must be a \texttt{<<reversed>>} port (required) and the other (the destination) must be a normal port (provided).
\end{wfr}

\begin{wfr}
\label{wfr:dir3}
If a link is typed with an association, the direction of the association must be conform to the direction of the link (derived from the direction of the ports at the ends).
\end{wfr}

Rule~\ref{wfr:dir1} restricts the links that can be used and we summarize here which connectors are accepted in OMEGA2:
\begin{enumerate}
\item Part - Part link $\Rightarrow$ assembly link, needs to be typed with an association\footnote{The need of typing a link with an association is given by the fact that a component has to know how to address the connector (see Rule~\ref{wfr:st2} later on).}
\item Port - Port link
\begin{enumerate}
\item {One port owned by the composite structure, the other one owned by a part\\
         port required - port required $\Rightarrow$ outbound delegation link\\         
         port provided - port required $\Rightarrow$ forbidden\\         
         port required - port provided $\Rightarrow$ forbidden\\         
         port provided - port provided $\Rightarrow$ inbound delegation link}
\item {Both ports are owned by parts\\
         port required - port required $\Rightarrow$ forbidden\\         
         port provided - port required $\Rightarrow$ assembly link\\         
         port required - port provided $\Rightarrow$ assembly link\\        
         port provided - port provided $\Rightarrow$ forbidden}
\end{enumerate}
\item Part - Port link
\begin{enumerate}
\item {Port owned by a part\\
         part - port provided $\Rightarrow$ assembly link, needs to be typed with an association\\      
         part - port required $\Rightarrow$ assembly link}
\item {Port owned by the composite structure\\
         part - port provided $\Rightarrow$ inbound delegation link\\         
         part - port required $\Rightarrow$ outbound delegation link, needs to be typed with an association}
\end{enumerate}
\end{enumerate}

The third rule introduces more constraints in the profile by establishing a correspondence between the direction of a connector typed with an association and the related association. These types of connectors need to be treated carefully so that by typing a link with an association, the direction in which it transports the messages does not become inconsistent and therefore the composite structure is not well-formed. This rule can be expanded in three cases:
\begin{itemize}
  \item The association is navigable at both ends. This type of association is accepted only for a link that connects two parts and the types of each end of the link and association must be compatible.
  \item The association is navigable only at one end. Then the types at each end of the link and the association should be compatible. This restricts us the associations that we may have:
    \begin{itemize}
	    \item For a link between two parts, the accepted associations are associations between two classes, a class and an interface or two interfaces. 
	    \item For a link from a part to a port, the accepted associations are between a class and an interface pointing to the interface or between two interfaces. 
	    \item For a link from a port to a part in this direction, only the association between two interfaces is accepted.
	    \item For a link between two ports, only the association between two interfaces is accepted.
    \end{itemize}
\item The association is not navigable at both ends then the connector is not well-formed.
\end{itemize}
The end of the link is compatible with the corresponding end of the association means:
\begin{itemize}
\item If the end of the link is a part and the association end is a class then the association end has to be equal or a supertype for the link's part type.
\item If the end of the link is a part and the association end is an interface then link's end type has to realize directly or indirectly the association's end type.
\item If the end of the link is a port and the association end is an interface then the port has to provide/require the association end.
\end{itemize}

We have mentioned as notion the \textit{direction of a link}. We can establish the direction based on a link's type and we can define the following. A connector starts from a port providing interfaces if:
\begin{itemize}
\item It is an inbound delegation link between provided ports and the port is owned by the composite structure;
\item It is an inbound delegation link between a provided port and a part.
\end{itemize} A connector starts from a port requiring interfaces if:
\begin{itemize}
\item It is an outbound delegation link between required ports and the port is owned by the inner component;
\item It is an assembly link between provided-required ports;
\item It is an assembly link between a part and a required port.
\end{itemize} A connector starts from an inner component if:
\begin{itemize}
\item It is an assembly link between a part and a provided port;
\item It is an outbound delegation link between a part and a required port;
\item It is an assembly link between parts.
\end{itemize}

\begin{figure}[!t]
\begin{center}
\includegraphics[width=10cm]{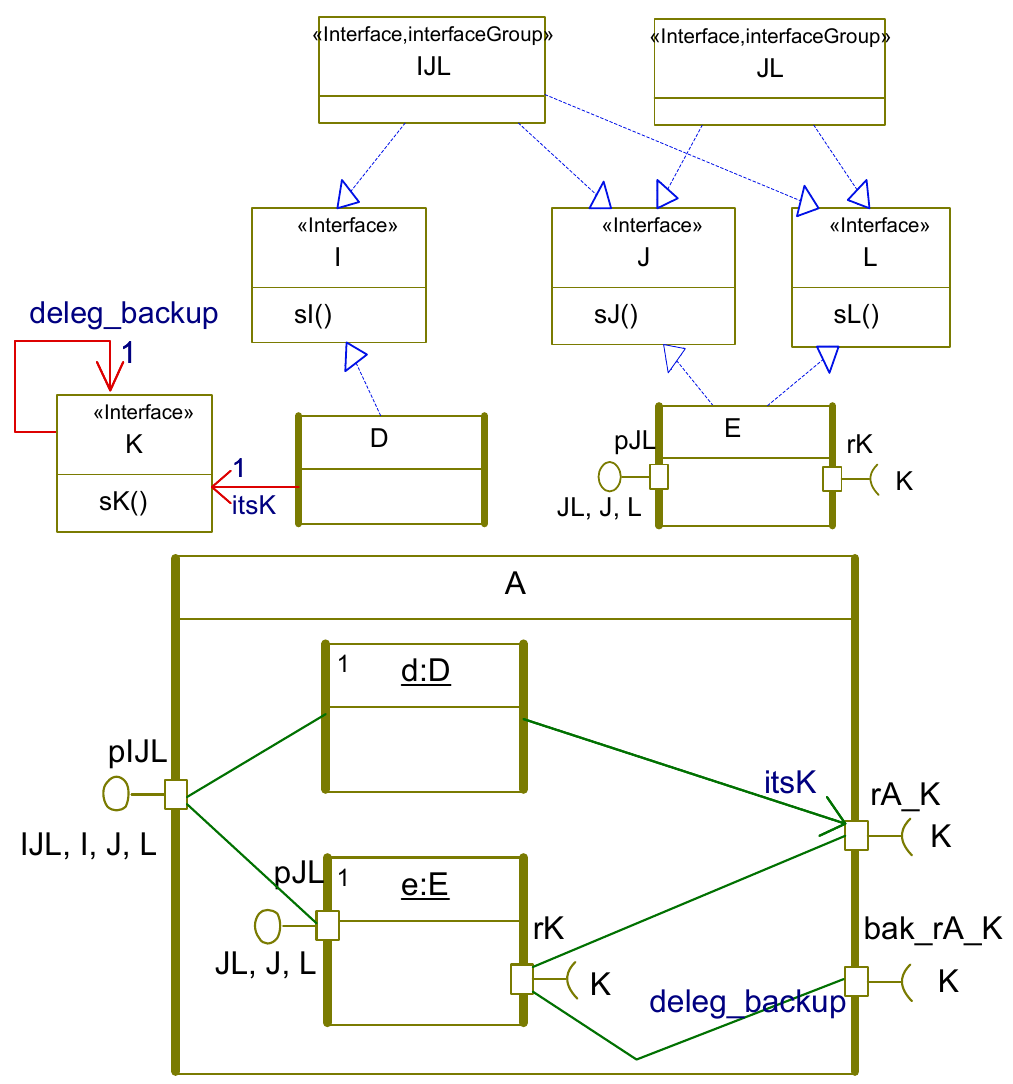}
\caption{Connection rules in composite structures}
\label{fig:delegEx}
\end{center}
\end{figure}

Taking as a running example the Figure~\ref{fig:delegEx} we can express the reason behind these rules. The example is a composite $A$ with two sub-components of types $D$ and $E$, one using ports for communication ($E$) and one not ($D$). For both sub-components there are incoming links (links from port $pIJL$ of $A$) and outgoing links (links to ports $rK$ and $bak\_rA\_K$ of $A$).

So, Rule~\ref{wfr:dir1} forbids putting a connector, for example between $pIJL$ and $rK$, since the direction of the connector would be ambiguous. Rule~\ref{wfr:dir3} forces the direction of a connector to be coherent with the direction of the realized association, like in the case of the link between $d$ and $rA\_K$ (realizing association $itsK$). 

These three rules allow us to have an overview on the composite structure and its behaviour. We can follow the flow of requests based on link's direction and establish which component reacts such that the main goal is achieved.

\subsection{Type coherence rules}
\label{sec:TypeCohR}
Before presenting the type system for connectors and type-based rules, we need to introduce some notions: \textit{interface groups}, \textit{default delegation associations} and \textit{set of transported interfaces}.

\paragraph{Interface groups.} Let us note that it is sometimes necessary to declare several provided or required interfaces for one port (for example, $pIJL$ of $A$ which provides interfaces $I$, $J$ and $L$, see Figure~\ref{fig:delegEx}). In UML, this is done by declaring a new interface that inherits from these interfaces and by using this new interface as the port type ($IJL$ in Figure~\ref{fig:delegEx}). However, such interfaces are artificial syntactic additions to the model, and they should not be taken into consideration by the link compatibility rules stated in the following. In our example, $d$ and $e$ only realize interfaces $I$ and respectively $J$ and $L$, so interface $IJL$ is irrelevant for the semantics of the model. In OMEGA2, such interfaces must be stereotyped with \texttt{<<interfaceGroup>>} to distinguish them from meaningful ones, as shown in the upper part of Figure~\ref{fig:delegEx}.

\paragraph{Default delegation associations.} The default behaviour of a port is to forward requests from one side to the other according to its direction: to the environment, if it is a \textit{required} port, and to its owner, if it is a \textit{provided} port. The minimum information needed by the port is, for each provided/required interface, which the destination should be. For example in Figure~\ref{fig:delegEx}, port $pIJL$ needs to know (and be able to refer to) the destination of requests belonging to interface $I$ (here, $d$) and the destination of requests belonging to $J$ or $L$ (here, $pJL$ of $e$). Similarly, $rK$ needs to know that the destination of outgoing requests is by default $rA\_K$.

It follows that, for each provided/required interface, the port has to possess an association designating to which the port should forward requests belonging to that interface. In OMEGA2, every interface type $I$ has by default an association called $deleg\_I$ pointing to itself, used for this purpose (for modelling convenience, the semantics considers they exist by default if they are omitted in the model). These associations are used to define the forwarding semantics of ports, described later on.

\paragraph{The dynamic type of a connector.} The type of a connector determines what type of invocations (signals or operation calls) can travel through the connector and how port behaviour descriptions refer to the connector. In general, in the case of a connector \textit{originating}\footnote{According to link directionality, as explained in Section~\ref{sec:DirR}.} from a port (i.e., not directly from a part), its type can be derived from the type of the entities situated at its two ends and does not necessarily need to be statically specified using an \textit{association}. The following notion defines the dynamic type of the connector:

\textbf{Definition 1 [Set of transported interfaces]}. For a connector starting from a port, the \textit{set of transported interfaces} is defined as the \emph{intersection} between the two sets of interfaces provided/required at the two ends of the link.

As the ends of a link can be either ports or components, the meaning of provided/required interfaces is defined for each case:
\begin{itemize}
\item For a \textit{Port}, the set of required/provided interfaces is the set containing the \textit{Port}'s type and all its supertypes, without all the interfaces stereotyped as \texttt{<<interfaceGroup>>}.

\item For a component, the set of provided interfaces is the set of all interfaces directly or indirectly realized by the component's class. 
\end{itemize}

According to this definition, the set of transported interfaces for the links in Figure~\ref{fig:delegEx} are as follows\footnote{Link $d$ to $rA\_K$ starts from the part $d$ and therefore the set of transported interfaces cannot be computed; moreover the link has to be statically typed (see Rule~\ref{wfr:st2} later on).}:
\begin{itemize}
\item For link $pIJL$ to $d$ the set is $\{I\}$.
\item For link $pIJL$ to $pJL$ the set is $\{J,L\}$.
\item For link $rK$ to $rA\_K$ the set is $\{K\}$.
\end{itemize}

Let us note that the link from $pIJL$ to $d$ given as example above could have been statically typed with association $deleg\_I$, because the set of transported interfaces $\{I\}$ only contains one element. However, in the general case when the derived set contains several interfaces (like for example the link between $pIJL$ and $pJL$ which transports $\{J,L\}$), statically typing a link with an association is not necessary and may be restrictive.

If a static type is specified, it must be compatible with the dynamic type, as stated in the following rule:

\begin{wfr}
\label{wfr:st1}
If a link outgoing from a port is statically typed with an association, then the association is necessarily directed (cf. Rule~\ref{wfr:dir3}) and the type pointed at by the association must belong to the \emph{set of transported interfaces} for that link.
\end{wfr}

On the example in Figure~\ref{fig:delegEx}, Rule~\ref{wfr:st1} implies that, for example, if the link $pIJL$ to $pJL$ is statically typed with an association then the association must point at either $J$ or $L$. But this restricts the set of requests forwarded through the link to only those requests which belong to the pointed interface ($J$ or $L$), therefore the behaviour is restricted compared to a dynamically typed link.

While the type for a connector starting \textit{from a port} does not need to be statically specified as it can be derived as shown before, if the connector starts directly \textit{from a component} (and not from a port) then the static type \textit{must} be specified: 

\begin{wfr}
\label{wfr:st2}
If a link originates in a component, then the link \emph{must} be statically typed with an association, and the type of the entity at the other end of the link must be compatible  with (i.e. be equal or a subtype of) the type at the other end of the association.
\end{wfr}

In Figure~\ref{fig:delegEx}, only the link from $d$ to $rA\_K$ is in this case; the link has indeed to be typed (here, with $itsK$) or otherwise the component would have no means to refer to it for communication.

Finally, a link is meaningful only if it can transport some requests:

\begin{wfr}
\label{wfr:nonvoid}
The set of transported interfaces for each link should not be void.
\end{wfr}

The above rules allow us to specify exactly what requests (signals and operation calls) can travel through connectors by defining compatible interfaces for each component. 

\subsection{Port behaviour rules}
\label{sec:portBeh}

\begin{figure}[!h]
\begin{center}
\includegraphics[width=7cm]{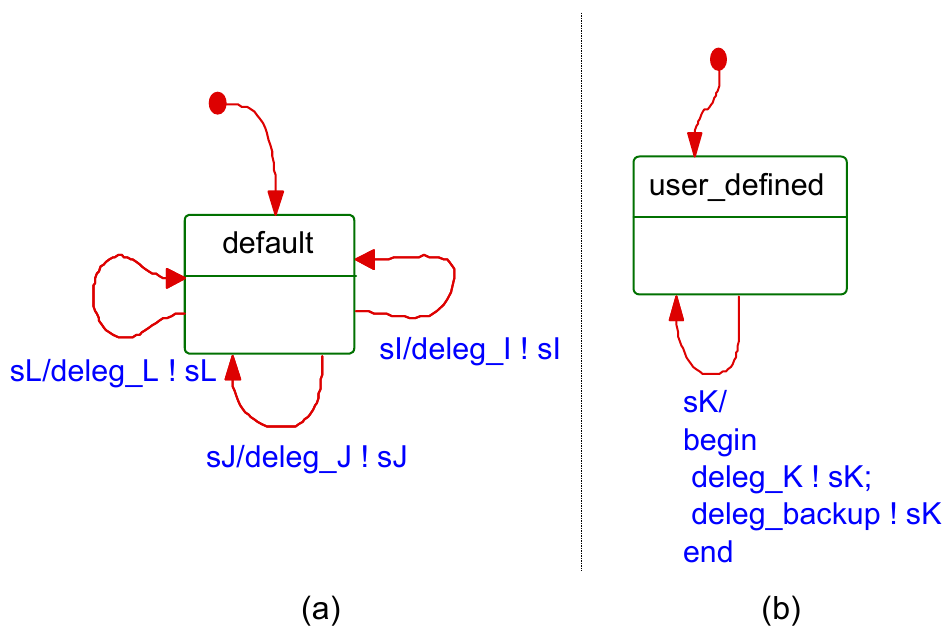}
\caption{(a) - Default state machine for port $pIJL$, (b) - User-defined machine for port $rK$}
\label{fig:portSM}
\end{center}
\end{figure}

In OMEGA2, the default behaviour of a port is to forward requests from one side to the other, depending on the port's direction. Each request (signal or operation call) will be forwarded to a destination which depends on the interface to which the signal or operation belongs, using the default $deleg$ associations above described. For example, the default forwarding behaviour of port $pIJL$ from Figure~\ref{fig:delegEx} can be described by the state machine in Figure~\ref{fig:portSM}-a\footnote{\texttt{deleg\_I!sI} is the OMEGA2 syntax for the action of sending signal \texttt{sI} to the destination \texttt{deleg\_I} (if the signal has formal parameters and no actual parameters are specified in the sending action, the actual values that will be sent are those ones received at the last \textit{reception} -- here the one that triggered the  transition).}.

The default behaviour is unambiguous only if for any interface, the entity to which the corresponding $deleg$ association points at is clear. Therefore, the following rules are necessary:

\begin{wfr}
\label{wfr:st3}
If several \emph{non-typed} connectors start from one port, then the sets of interfaces transported by each of the connectors have to be pairwise disjoint.
\end{wfr}

The last rule does not forbid the case where a port is connected to $n$ entities that provide or require the same interface $I$ ($n>1$): it states that in this case at least $n-1$ connectors have to be explicitly typed with associations. The one connector which is not explicitly typed, if it exists, is implicitly typed with $deleg\_I$. In the example from Figure~\ref{fig:delegEx}, port $rK$ of $e$ is in this situation: it has two links to two ports ($rA\_K$ and $bak\_rA\_K$), both typed with the same interface ($K$). According to Rule~\ref{wfr:st3}, one of the links has to be explicitly typed; here, the second one is statically typed with the association $deleg\_backup$.

The default port behaviour may be redefined by attaching a state machine to the port's type. In OMEGA2, this state machine may use the implicitly typed connectors ( accessed via the default $deleg$ associations), as well as the explicitly typed connectors (via their defining association). In Figure~\ref{fig:portSM}-b we show an example of port behaviour for port $rK$ (from Figure~\ref{fig:delegEx}), which duplicates every $sK$ signal on both the default connector ($deleg\_K$, communicating with $rA\_K$) and the secondary connector ($deleg\_backup$, communicating with $bak\_rA\_K$).

In addition, for completeness of the port behaviour, we require the following:
\begin{wfr}
\label{wfr:st4}
The union of the sets of interfaces transported by each of the connectors \emph{originating} from a port $P$ must be equal to the set of interfaces provided/required by $P$.
\end{wfr}

Applied for example to port $pIJL$ from Figure~\ref{fig:delegEx}, this rule says that the two links originating from the port must transport, together, the entire set of interfaces provided by the port, i.e. $\{I,J,L\}$ (remember that $IJL$ is an \texttt{<<interfaceGroup>>} and does not count in type checks).

\subsection{Concurrency model and observers}
\label{sec:ConcOM2}

The concurrency model is left open in UML. The previous version of OMEGA defined a particular concurrency model, based on the standard UML notion of \textit{active} and \textit{passive} classes. Due to the choice of partitioning the object space in activity groups attached to active objects, certain forms of simple resource sharing and synchronisation generated quite complex models, as sharing could only be achieved via an explicitly modelled resource manager -- an active object. In order to overcome this problem, a new kind of passive class can be defined in OMEGA2 (using the stereotype \texttt{<<protected>>}).

\textit{Protected objects} are passive objects that do not belong to one activity group but rather are shared between the groups. They work in the same way as Ada protected objects \cite{Ada2005}. Like in Ada, protected objects are a synchronization mechanism. They provide \textit{functions} (which may only read but not modify object attributes) that can be executed concurrently, and \textit{entries} that are executed in mutual exclusion from each other and from functions (this corresponds to the classical readers-writers pattern). In addition, an entry has a guard; a call to an entry from a thread (activity group) will wait before beginning the execution until the guard is true. Our model of protected objects is slightly simplified (more non-deterministic) compared to the \textit{eggshell} model of Ada \cite{Ada2005} and therefore suppresses the need for \textit{procedures} existing in Ada: a procedure can be seen as an entry with guard \texttt{true}.

The OMEGA2 concurrency model therefore distinguishes three kinds of classes: active, passive and protected. Since every passive object is considered to \textit{belong} to an active object, in the sense that its behaviour is executed on the execution thread of its owner, some rules are necessary to avoid confusing configurations in composite structures.

\begin{figure}[!h]
\begin{center}
\includegraphics[width=8cm]{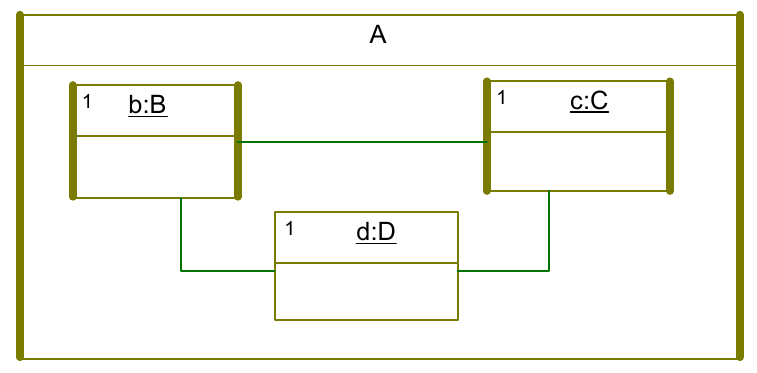}
\caption{Forbidden composite structure.}
\label{fig:forbiddenStr}
\end{center}
\end{figure}

For example, the composite structure in Figure~ \ref{fig:forbiddenStr} shows, on the same level, two active objects ($b$ and $c$), which have their own \textit{activity group}, and a passive object ($d$) which belongs to the group of its creator (instance of $A$). This kind of structure is forbidden. If the desired semantics is to have a shared passive object $d$, then $d$ may be declared as \texttt{<<protected>>} and the structure becomes valid.

\begin{wfr}
\label{wfr:comp1}
A \textit{passive} class may define a composite structure formed only of passive classes.
\end{wfr}

\begin{wfr}
\label{wfr:comp2}
An \textit{active} class may define a composite structure formed of either only passive classes, or of a combination of active and protected classes.
\end{wfr}

As an extension to the original profile, an observer can also define a simple composite structure. Composite observers have proved to be a way for making more compact the specification of some complex verification properties. The well-formedness rule is:

\begin{wfr}
If an \textit{Observer} defines a composite structure, the components must also be instances of \textit{Observers}.
\end{wfr}

\section{Translation to IF}
\label{ch:CompStruct2IF}

The mapping between the OMEGA Profile and the IF language is based on the principles explained in this section. 

Every UML class is mapped to a process with a local variable for each attribute or association of the class. Inheritance between classes is translated by the duplication of each inherited attribute in the processes corresponding to subclasses. Operations are defined by signals, while statemachines are signals are translated almost syntactically to IF. For the time extension \textit{clocks} exist as a predefined type in IF and \textit{timers} are translated using a clock and a timer process sending \textit{timeout} signals. For further details the reader is referred to \cite{OberGO06}.

The translation of composite structures is based on the principle that the modelling elements involved in composite structures, namely ports and connectors, should be handled as first class language citizens. This means that we refrain from flattening the model during compilation and hard-wiring all the communication paths (something that is done, for example, in certain SDL compilers). Concretely, each port instance is implemented as an IF process instance (whose behaviour corresponds to the routing behaviour described in Section~\ref{sec:portBeh}) and each connector is represented by attributes in the end-points (in ports or in components), corresponding to the association defining the connector (the default $deleg$ association or the explicitly specified one).

In this setting, a UML composite structure diagram is used simply as an initialization scheme for instantiating components and ports and for creating links. A composite structure  is therefore translated to a constructor.

As a consequence of the translation sketched above, a signal or operation call sent through a connector chain will pass through several objects (the intermediate ports) before reaching the destination. In order to avoid the state space explosion problem due to the interleaving of such ``forwarding'' actions, the translator defines a \textit{total priority order} between these actions. Thus, even if several signals are in transit on connector chains, only one forwarding action (belonging to the enabled port with the highest priority) will be enabled at any given time. This yields an increase of state space due to connector chains which is linear in the length of the chain, instead of combinatorial explosion. Note that starvation of lower priority actions is not possible since, in any state, eventually all signals that are in transit through connectors will arrive at destination and the rest of the system will be able to make progress. Moreover, this abstraction is made without any loss of generality, since all the possible interleavings at the level of component transitions (which is the observable level) remain feasible. The implementation of the abstraction is made very easy by the dynamic priority mechanism of IF .

Another element that is added in the second version of the OMEGA profile is \textit{protected} classes. Compared to normal passive classes, protected classes add the classical \textit{readers-writers} synchronization protocol for functions and entries. The readers-writers protocol implemented in our translation is a variant of the classical solution that may be found in many textbooks (e.g., \cite{BurnsWellings01}). The implementation is however facilitated by the fact that the IF language offers mutually exclusive and atomic \textit{transitions} by default, and transitions can specify conditional waiting simply using guard conditions.

For structured observers, the same mapping as for composite structures is applied.

\section{OCL Formalization}
\label{ch:OCLF}

The rules disambiguating composite structures also implemented in the OMEGA2 compiler are formalized in OCL to verify that UML models comply with our profile. The OCL code was developed in Topcased OCL Environment \cite{topcased-web}.

For our formalisation we have defined helper functions for accessing:
\begin{itemize}
\item type of elements connected with a link: \textit{has2Ports}, \textit{has2Parts}, \textit{has1PartAnd1Port}, \textit{has1PartWithPort}, \textit{has2PartsWithPort};
\item the connected elements: \textit{part1}, \textit{part2}, \textit{port1}, \textit{port2};
\item association's properties: \textit{isTyped}, \textit{isBidirectional}, \textit{isNotNavigable}, \textit{isEnd1Navigable}, \textit{associationEnds}, \textit{isClassClassAssociation}, \textit{isInterfaceInterfaceAssociation}, \textit{associationStartPoint}, \textit{associationStartPointType}, \textit{associationEndPoint}, \textit{associationEndPointType}, \textit{isClassInterfaceAssociation};
\item port's type: \textit{isReversed};
\item types of classifiers: \textit{isInterfaceGroup}, \textit{isInterface}, \textit{isProtected}, \textit{isObserver}.
\end{itemize}
The definition of these functions, together with the invariants presented in this Section, can be found in the Appendix.

For the formalization of Rule~\ref{wfr:dir1} and Rule~\ref{wfr:dir2} we define a function that will compute the exact link type based on the classification presented in Section~\ref{sec:DirR}. The OCL invariant corresponding to these two rules becomes the verification for each connector in the model if its type is not \textit{forbidden}:
\bs
\begin{lstlisting}[language=OCL, frame=single]
context Connector

-- Definition of link's type
def: linkType : String = 
  if has2Parts 
    then 'assembly link between parts'
  else 
    if has2Ports 
      then if has1PartWithPort
             then if not port1.isReversed and not port2.isReversed 
                    then 'inbound delegation link between provided ports'
                  else if port1.isReversed and port2.isReversed
                         then 'outbound delegation between required ports'
                       else 'forbidden'
                       endif
                  endif
           else if (port1.isReversed and port2.isReversed) or
                   (not port1.isReversed and not port2.isReversed)
                  then 'forbidden'
                else  'assembly between provided-required ports'           
                endif
           endif
    else  
      if has1PartWithPort
        then if not port1.isReversed
               then 'assembly link between part and provided port'
             else 'assembly link between part and required port'
             endif
      else if not port1.isReversed
             then 'inbound delegation link between part and provided port'
           else 'outbound delegation link between part and required port'
           endif
      endif
    endif        
  endif

-- Rule 1 and Rule 2
inv LinkType: self.linkType <> 'forbidden'
\end{lstlisting}
\bs

For the formalization of the third rule we need to verify the compatibility between the association end (of the association typing the link) and the corresponding link end (i.e. the compatibility has to be verified between the start point for both link and association and for the end point). As explained in Section~\ref{sec:DirR}, this resumes to verify the inclusion of the association's end type in link's end type (realized interfaces or superclasses). We define functions that verify if a link starts from a port or a part as already presented (\textit{isStartingFromProvidedPort}, \textit{isStartingFromRequiredPort}, \textit{isStartingFromPort}, \textit{isStartingFromPart}), and for each link which is its starting point and its ending point (\textit{linkStartPort}, \textit{linkEndPort}, \textit{linkStartPart}, \textit{linkEndPart}). Since the formalization of the following rules consists in computing the provided/required interfaces for a port and a component and since the same calculus can be used for Rule~\ref{wfr:dir3}, we shall continue by computing the needed sets. 

We continue with the calculus of the provided/required interfaces for a port and the interfaces provided by a component. In the case of a port, the set of realized interfaces is given by the set of provided interfaces without those stereotyped with \verb+<<interfaceGroup>>+.\footnote{Required interfaces are modelled with reversed ports and are therefore also accessed using \texttt{provided}.} Please note that interfaces stereotyped \verb+<<interfaceGroup>>+ are artificially added to the model and they should not be taken into consideration in our formalization.
\bs
\begin{lstlisting}[language=OCL, frame=single]
context Port

-- Definition of interfaces realized by a port
def: interfaces : Set(Classifier) = self.provided->reject(isInterfaceGroup)
\end{lstlisting}

\bs
Before computing the set of realized interfaces by a class (the type of a part), we have to compute recursively the list of parents. We suppose that our model is well-formed and has no cycles.
\bs
\begin{lstlisting}[language=OCL, frame=single]
context Classifier

-- Definition of classifier's parents recursive computation
def: getParentsRec : Set(Classifier) =  self.general->union(self.general->iterate(p:Classifier; res:Set(Classifier)=Set{}| res->union(p.getParentsRec)))
\end{lstlisting}

\bs
For a class, the set of provided interfaces is the set of realized interfaces summed with the set of parents for each realized interface and summed with the set of provided interfaces for each parent of our class, without those stereotyped with \verb+<<interfaceGroup>>+.
\bs
\begin{lstlisting}[language=OCL, frame=single]
context Class

-- Definition of all interfaces directly realized by a class
def: iRealizations : Set(Classifier) = self.interfaceRealization.contract->asSet()

-- Definition of interfaces provided by a class directly or indirectly realized (used in the case of a link not typed by an association)
def: interfaces : Set(Classifier) = 
  iRealizations->union(iRealizations->iterate(i:Interface; res:Set(Classifier)=Set{}| res->union(i.getParentsRec)))
  ->union(self.getParentsRec->iterate(c:Class; res:Set(Classifier)=Set{}| res->union(c.interfaces)))
  ->reject(isInterfaceGroup)
\end{lstlisting}

\bs
In the case of a link typed with an association which has as an end an interface, the set of provided interfaces is the set of the interface to which it points summed with its parents and without those interfaces stereotyped with \verb+<<interfaceGroup>>+.
\bs
\begin{lstlisting}[language=OCL, frame=single]
context Interface

-- Definition of interfaces provided by an interface (used in the case of a link typed by an association pointing to an interface)
def: interfaces : Set(Classifier) = self.oclAsType(uml::Classifier)->asSet()->union(self.getParentsRec)->select(not isInterfaceGroup and isInterface)
\end{lstlisting}

\bs
Because of the mishandling of polymorphic functions in OCL, we need to define explicitly the polymorphism of the function \textit{interfaces} on the subtypes of \textit{Type} (\textit{Class} and \textit{Interface}).
\bs
\begin{lstlisting}[language=OCL, frame=single]
context Type

-- Determines the set of provided interfaces by a class or an interface
def: interfaces : Set(Classifier) =
  if self.oclIsKindOf(uml::Interface)
    then self.oclAsType(uml::Interface).interfaces
  else
    self.oclAsType(uml::Class).interfaces
  endif
\end{lstlisting}

\bs
We compute the set of transported interfaces as the intersection of provided interfaces of both ends and we formalize Rule~\ref{wfr:nonvoid}: the cardinal of the set of transported interfaces should be at least equal to one. We need to remark that this  set is computed for links starting from a port (typed or not typed with an association) and it is not computed in the case of a part-part connector. 
\bs
\begin{lstlisting}[language=OCL, frame=single]
context Connector

-- Definition of the set of transported interfaces
def: setTransportedInterfaces : Set(Classifier) =
  if has2Parts
    then Set{OclInvalid}
  else if has2Ports
         then if isTyped
                then if isStartingFromPort(port1)
                       then (port1.interfaces) -> intersection(associationEndPointType.interfaces)
                     else (port2.interfaces) -> intersection(associationEndPointType.interfaces)
                     endif
              else (port1.interfaces) -> intersection(port2.interfaces) 
              endif                              
       else  
         if isTyped
           then (port1.interfaces) -> intersection(associationEndPointType.interfaces)
         else (port1.interfaces) -> intersection(part1.type.interfaces)      
         endif 
       endif
  endif        

-- Rule 6  
inv SetOfTransportedInterfacesNonEmpty: self.setTransportedInterfaces->size()<>0
\end{lstlisting}

\bs
In order to formalize Rule~\ref{wfr:st1} and Rule~\ref{wfr:st2}, we define the compatibility between two classes and between a class and an interface as the inclusion of association's end type in the set of provided interfaces or in the set of parents. It is followed by the compatibility between a port and an interface as the inclusion of all realized interfaces by the association's end type in the set of provided/required interfaces of the port. 
\bs
\begin{lstlisting}[language=OCL, frame=single]
context Classifier

-- Verifies if the current classifier is compatible with the one given as parameter
def: isCompatible(c:Classifier) : Boolean =
  if self.oclIsKindOf(uml::Interface) and c.oclIsKindOf(uml::Interface)
    then self.oclAsType(uml::Interface).interfaces->includes(c)
  else
    if self.oclIsKindOf(uml::Class) and c.oclIsKindOf(uml::Interface)
      then self.oclAsType(uml::Class).interfaces->includes(c)
    else
     (c->asSet()->union(c.getParentsRec))->includes(self)
    endif
  endif
  
context Type

-- Verifies if the link end's type (the type of a part) is compatible with the association end's type given as parameter
def: isCompatible(t:Type) : Boolean = self.oclAsType(uml::Classifier).isCompatible(t.oclAsType(uml::Classifier))

context Port

-- Verifies if the current port is compatible with the association end's type given as parameter
def: isCompatible(t:Type) : Boolean = self.interfaces->includesAll(t.oclAsType(uml::Interface).interfaces)
\end{lstlisting}

\bs
Rule~\ref{wfr:st1} states that for a link starting from a port and typed with an association, the association must be directed (unidirectional) and the interface pointed by the association has to be included in the set of transported interfaces. We include here Rule~\ref{wfr:dir3}, which adds that the direction of the link has to be conforming to the direction of the association, as defined in Section~\ref{sec:DirR}. We define a function (\textit{linkStartingFromPortVerification}) that verifies if for a link typed with an interface-interface association (the only association accepted for a connector starting from a port) the corresponding start points and end points are compatible and that the interface to which it points is included in the set of transported interfaces.
\bs
\begin{lstlisting}[language=OCL, frame=single]
context Connector

-- Verifies if a link starting from a port and typed with an association has the same direction with the association and the interface pointed is included in the set of transported interfaces
def: linkStartingFromPortVerification : Boolean =
  if isNotNavigable or isBidirectional
    then false
  else 
    if isInterfaceInterfaceAssociation
        then if has1PartAnd1Port
              then linkStartPort.isCompatible(associationStartPointType) and linkEndPart.type.isCompatible(associationEndPointType) and 
                   setTransportedInterfaces->includes(associationEndPointType.oclAsType(uml::Classifier))
             else linkStartPort.isCompatible(associationStartPointType) and linkEndPort.isCompatible(associationEndPointType) and 
                  setTransportedInterfaces->includes(associationEndPointType.oclAsType(uml::Classifier))
             endif
    else false
    endif
  endif
\end{lstlisting}

\bs
Then the OCL invariant corresponding to these two rules verifies that for each link in the model starting from a port and typed with an association the compatibility stated above is verified.
\bs
\begin{lstlisting}[language=OCL, frame=single]
context Connector

-- Verifies if a link starting from a port is well-formed
def: linkStartingFromPort : Boolean =
  if (not isStartingFromPart) and isTyped
    then linkStartingFromPortVerification
  else true
  endif

-- Rule 3 and Rule 4
inv LinkStartingFromPort: self.linkStartingFromPort
\end{lstlisting}

\bs
Rule~\ref{wfr:st2} completes Rule~\ref{wfr:dir3}, by adding that all connectors starting from a part have to be typed with an association and if the association is bidirectional (the only bidirectional association accepted is the association between two classes that may type only the link that connects two parts) it has to be compatible with the link in a direction. For a unidirectional association that types the link, we need to have the compatibility between the corresponding ends (link's start part with association's start point and link's end part/port with association's end point). This is expressed by the below functions (\textit{linkPartPartVerification}, \textit{linkPartPortVerification}), which make the difference between a link that connects two parts (which accepts all kinds of associations) and the link that connects a part with a port (which accepts only the association between two interfaces or between a class and an interface).
\bs
\begin{lstlisting}[language=OCL, frame=single]
context Connector

-- For a link between two parts verifies if the ends are compatible with the corresponding ends of the accepted association 
def: linkPartPartVerification : Boolean =
  if isNotNavigable 
    then false
  else
    if isBidirectional
      then (linkStartPart.type.isCompatible(associationStartPointType) and linkEndPart.type.isCompatible(associationEndPointType)) or
              (linkStartPart.type.isCompatible(associationEndPointType) and linkEndPart.type.isCompatible(associationStartPointType)) 
    else (linkStartPart.type.isCompatible(associationStartPointType) and linkEndPart.type.isCompatible(associationEndPointType))
    endif
  endif

-- For a link between a part and a port verifies if the ends are compatible with the corresponding ends of accepted association
def: linkPartPortVerification : Boolean =
  if isNotNavigable or isBidirectional
    then false
  else 
    if isClassClassAssociation
      then false
    else
      if isInterfaceInterfaceAssociation
        then (linkStartPart.type.isCompatible(associationStartPointType) and linkEndPort.isCompatible(associationEndPointType))
      else
        if isClassInterfaceAssociation
          then linkStartPart.type.isCompatible(associationStartPointType) and linkEndPort.isCompatible(associationEndPointType)
        else false
        endif
      endif
    endif
  endif
\end{lstlisting} 

\bs
The invariant for Rule~\ref{wfr:dir3} and Rule~\ref{wfr:st2} verifies that each connector in the model starting from a part is typed with an association and the direction of the association is compatible with the direction of the link:
\bs
\begin{lstlisting}[language=OCL, frame=single]
context Connector

-- Verifies if a link starting from part is typed with an association and if it is well-formed
def: linkStartingFromPart : Boolean =
  if isStartingFromPart
    then if has2Parts
           then isTyped and linkPartPartVerification 
         else isTyped and linkPartPortVerification
         endif
  else true
  endif

-- Rule 3 and Rule 5
inv LinkStartingFromPart: self.linkStartingFromPart 
\end{lstlisting} 

\bs
We formalize now the rules for port behaviour. The default behaviour is that the port forwards the requests received according to its direction: to the environment if it is a \textit{required port} and to the component that owns it if it is a \textit{provided port}. This means that the port knows how to respond to any received request and also which is the destination of the request.

The context in our formalization becomes the \textit{Port} and we define functions that give all the connectors (typed or not with an association) starting from the port (\textit{connectors}, \textit{connectorsNotTyped}, \textit{connectorsStartingFromPort}) and that verify if the port has connectors (typed or not with an association) starting from it (\textit{hasConnectors}, \textit{hasTypedConnectors}, \textit{isStartingPort}).

The relation behind the first rule concerning port's behaviour states that the sets \begin{math}A_1, A_2,..., A_n\end{math} are \textit{pairwise disjoint} if and only if \begin{math}\mathrm{card}(A_1 \cup A_2 \cup ... \cup A_n)=\mathrm{card}(A_1)+\mathrm{card}(A_2)+...+\mathrm{card}(A_n)\end{math}. The function \textit{unionSetForTransportedInterfacesOnLinks} computes the left hand side of the equality, and the function \textit{noOfTransportedInterfacesOnLinks} computes the right hand side of the expression. 
\bs
\begin{lstlisting}[language=OCL, frame=single]
context Port

-- Determines the union of the sets of transported interfaces on each link starting from the port
def: unionSetForTransportedInterfacesOnLinks(withType:Boolean) : Set(Classifier) =
  self.connectorsStartingFromPort(withType)->iterate(c:Connector; s:Set(Classifier)=Set{} | s->union(c.setTransportedInterfaces))

-- Determines the sum of the number of transported interfaces on each link starting from the port 
def: noOfTransportedInterfacesOnLinks(withType:Boolean) : Integer =
  self.connectorsStartingFromPort(withType)->iterate(c:Connector; i:Integer=0 | i + (c.setTransportedInterfaces->size()))
\end{lstlisting}

\bs
Then Rule~\ref{wfr:st3} becomes the verification of the equality stated above:
\bs
\begin{lstlisting}[language=OCL, frame=single]
context Port

-- Definition of pairwise disjoint sets of transported interfaces
context Port
def: isPairwiseDisjoint : Boolean = 
  if self.connectorsStartingFromPort(false)->size() >= 2
    then  self.unionSetForTransportedInterfacesOnLinks(false)->size() = self.noOfTransportedInterfacesOnLinks(false)
  else true
  endif

-- Rule 7
inv PairwiseDisjoint: self.isPairwiseDisjoint
\end{lstlisting}

\bs
For Rule~\ref{wfr:st4} we will use the union of the sets of transported interfaces computed above and we will test its equality with the set of provided/required interfaces by the port.
\bs
\begin{lstlisting}[language=OCL, frame=single]
context Port

-- Verifies if the union of sets of transported interfaces is equal to the interfaces provided/required
def: isComplete : Boolean =
  if isStartingPort
    then unionSetForTransportedInterfacesOnLinks(true) = self.interfaces
  else true
  endif  

-- Rule 8
inv Completeness: self.isComplete       
\end{lstlisting}

\bs
For the two rules concerning the execution model for composite structures we will reason on the number of active, passive and protected components given by the functions \textit{noOfComponents}, \textit{noOfActiveComponents}, \textit{noOfPassiveComponents}, \textit{noOfProtectedComponents} and \textit{isComposite}. 

Rule~\ref{wfr:comp1} says that if a composite structure is passive then it is well-formed if and only if the number of passive parts this owns is equal with the total number of parts. Rule~\ref{wfr:comp2} says that if a composite structure is active then it is well-formed if and only if or the number of passive parts is equal to the total number of parts or the sum between the number of active parts and the number of protected parts is equal to the total number of parts. 
\bs
\begin{lstlisting}[language=OCL, frame=single]
context Class

-- Definition of a well-formed class
def: isWellFormed : Boolean = 
  if self.isActive and isComposite 
    then  if noOfActiveComponents + noOfProtectedComponents = noOfComponents or 
             noOfPassiveComponents = noOfComponents
            then true
          else false
          endif
  else
    if (not self.isActive) and (not isProtected) and isComposite
      then if noOfPassiveComponents = noOfComponents
             then true
           else false
           endif
    else OclInvalid
    endif
  endif

-- Rule 9 and Rule 10
inv CompositeStructure: self.isWellFormed <> false
\end{lstlisting}

\bs
The last rule regarding the simple composite observers is also formalized with the means of the number of observer parts. This rule is equivalent to: the number of parts of a composite observer is equal to the number of observer parts (\textit{noOfObservers}) of the same composite structure.
\bs
\begin{lstlisting}[language=OCL, frame=single]
context Class

-- Definition of a well formed observer
def: isObserverWellFormed : Boolean =
  if self.isObserver and isComposite
    then noOfComponents = noOfObservers
  else true
  endif

-- Rule 11
inv CompositeObserver: self.isObserverWellFormed <> false
\end{lstlisting}

\section{Evaluation}
\label{ch:eval}
To validate the approach, we evaluated the rules on several complex models. The most complex example we used is a model of the solar wings deployment system of the ATV\footnote{Automated Transfer Vehicle of the International Space Station, \url{http://www.esa.int/atv}} provided by Astrium Space Transportation. The model features a 3-level hierarchical architecture with 37 classes (7 composite ones), 93 active objects at runtime and approximately 380 ports and 200 connectors. 

The OCL formalization was applied on the model in order to test model compliance with the OMEGA2 profile and to search for modelling errors. Since the original model had not been built for simulation or verification, the first issue pointed out by the rules was ports and connectors were untyped. The corrective action consisted in defining a total of 26 interfaces, and using them for specifying port contracts. Only a few ports in the original model were bidirectional and splitting them to unidirectional ports did not raise problems, resulting in a clearer model. The evaluation of the OCL rules yielded the inconsistent ports and connectors (cf. Section~\ref{sec:BidirPort} and Section~\ref{sec:DirR}) which were either removed or redefined. 

\begin{figure}[!ht]
\centering
\includegraphics[width=14cm]{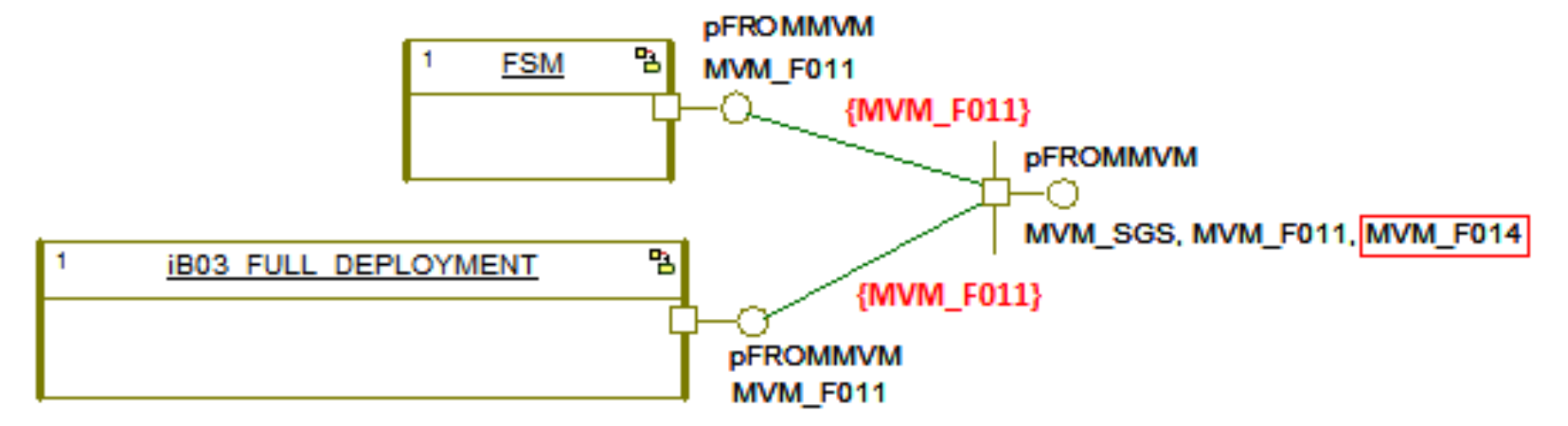}
\caption{Inconsistent port with respect to uniqueness and completeness rules}
\label{fig:inconsistentPort}
%\end{center}
\end{figure}

A second task was the verification of the uniqueness and completeness of ports, (cf. Section~\ref{sec:TypeCohR}-Section~\ref{sec:portBeh}). Approximately 20\% of the evaluated ports were inconsistent with respect to rules~\ref{wfr:st3} and ~\ref{wfr:st4}. Figure~\ref{fig:inconsistentPort} shows one such example.

Finally, the corrected model was given as input to the OMEGA2 compiler and was simulated with the IFx2 toolset. During simulation, deadlocks due to missing connectors or unhandled requests by ports were not found. Given the complexity of the models, this provides strong empirical evidence that, under the constraints of the rule set, the OMEGA2 type system is safe.

\section{Conclusions}
\label{ch:concl}

Composite structures play an important role in modelling real-time embedded systems. They offer a clear structure of these systems and an initialization scheme for the objects contained. They are a big evolution of the UML standard version 2.x, since in the version 1.4 the initialization order of complex systems was user-defined. Since the standard is ambiguous and semantic variation point left open, we proposed to define a rule set observing composite structure and to prove its type safety.

We presented a definition and formalization of an operational model of UML composite structures, our approach being based on :
\begin{itemize}
\item dynamic typing of connectors based on a derived notion of \textit{transported interfaces};
\item a set of static well-formedness rules, including type checking rules;
\item a full definition of the default behaviour of \textit{Ports}, and the means for defining port behaviour differing from the default (by using implicit port associations, etc.)
\item rules for relating composite structures with the concurrency model.
\end{itemize}

The rule set defined in Chapter~\ref{ch:OCLF} is used by the type checker of the OMEGA UML compiler. In addition, the compiler goes all the way down to an operational implementation of composite structures, by translating OMEGA UML models (edited with any XMI 2.0 compatible UML editor) into IF models, for which a simulation and model-checking platform exists allowing us to prove the correctness of UML embedded models. 

Experiments have been conducted to prove that models observing this rule set are correct. While the OMEGA UML compiler is able to catch all modelling errors when translating the model into its IF description, the OCL formalisation can also reveal these issues in a step preceding the translation. Applying this formalisation on the model, it yields the elements that do not comply with our profile catching many corner cases.

The next step in our work is to prove the type-safety of our ruleset with respect to composite structures using the Isabelle/HOL proof assistant \cite{Paulson94}. In this setting the type-safety means that: any request that travels through connectors reaches its terminus and every destination object receives only request compatible with its interfaces. Even thought we were able to show on realistic models using the simulation and exhaustive state-space search from IFx2 Toolset that no routing problems (deadlocks in ports due to missing links, unexpected requests not conforming to object interfaces, etc.) exist in the model, a formal proof is needed. 

\nocite*
\bibliographystyle{plain}
\bibliography{Well-formedness_and_typing_rules_for_UML_Composite_Structures}

\newpage
\appendix
\section*{Appendix: OCL Formalization}
\markboth{APPENDIX}{APPENDIX}
\addcontentsline{toc}{section}{\protect\numberline{}Appendix} 
\begin{lstlisting}[language=OCL, caption=]
-- HELPER FUNCTIONS

context Classifier

-- Verifies if the classifier is an interface
def: isInterface : Boolean = self.oclIsTypeOf(uml::Interface)

-- Verifies if the classifier is stereotyped with <<interfaceGroup>>
def: isInterfaceGroup : Boolean = self.getAppliedStereotypes()->select(name='interfaceGroup')->size()<>0

-- Verifies if the classifier is stereotyped with <<protected>>
def: isProtected : Boolean = self.getAppliedStereotypes()->select(name='protected')->size()<>0

--Verifies if the classifier is stereotyped with <<observer>>
def: isObserver : Boolean = self.getAppliedStereotypes()->select(name='observer')->size()<>0


context Port

-- Verifies if the port is reversed (the port requires interfaces)
-- def: isReversed : Boolean = self.getAppliedStereotypes()->select(name='reversed')->size()<>0

-- Since the tool used for the development of our models is IBM Rhapsody tool we need to make some adjustments since Rhapsody supports the reversed mechanism and saves it as an attribute of RhpPort stereotype 
def: isReversed : Boolean = self.getValue(self.getAppliedStereotypes()->select(name='RhpPort')->asOrderedSet()->at(1),'isReversed').oclAsType(Boolean)


context Connector

-- Verifies if the connector is of type port-port
def: has2Ports : Boolean = self.end.role->select(oclIsTypeOf(uml::Port))->size() = 2

-- Verifies if the connector is of type part-part
def: has2Parts : Boolean = self.end.role->select(oclIsTypeOf(uml::Property) and not oclIsTypeOf(uml::Port))->size() = 2 

-- Verifies if the connector is of type port-part
def: has1PartAnd1Port : Boolean = self.end.role->select(oclIsTypeOf(uml::Property) and not oclIsTypeOf(uml::Port))->size() = 1

-- For a port-port connector verifies if one of the ports is owned by an inner component and the other one by the composite structure
-- For a port-part connector verifies if the port is owned by an inner component
def: has1PartWithPort : Boolean = self.end.partWithPort -> reject(oclIsTypeOf(OclVoid))->size() = 1

-- For a port-port connector verifies if both ports are owned by inner components
def: has2PartsWithPort: Boolean = self.end.partWithPort -> reject(oclIsTypeOf(OclVoid))->size() = 2

-- For a port-port connector returns the first port from the set of ends
-- For a port-part connector returns the port
def: port1 : Port = self.end.role->select(oclIsTypeOf(uml::Port))->asOrderedSet()->at(1).oclAsType(uml::Port)

-- For a port-port connector returns the second port from the set of ends
def: port2 : Port = self.end.role->select(oclIsTypeOf(uml::Port))->asOrderedSet()->at(2).oclAsType(uml::Port)

-- For a part-part connector returns the first part from the set of ends
-- For a port-part connector returns the part
def: part1 : Property = self.end.role->select(oclIsTypeOf(uml::Property) and not oclIsTypeOf(uml::Port))->asOrderedSet()->at(1).oclAsType(uml::Property)

-- For a part-part connector returns the second part from the set of ends
def: part2 : Property = self.end.role->select(oclIsTypeOf(uml::Property) and not oclIsTypeOf(uml::Port))->asOrderedSet()->at(2).oclAsType(uml::Property)

-- Verifies if the connector is typed with an association
def: isTyped : Boolean = (not self.type.oclIsTypeOf(OclVoid)) 

-- For a part-part connector, in the case of a unidirectional association that may type the link we need to know which end is navigable such that we can determine the direction of the link
def: isEnd1Navigable : Boolean = self.end.definingEnd->asOrderedSet()->at(1).isNavigable()

-- Determines the ends of the association with which the link is typed
def: associationEnds : OrderedSet(Property) = self.type.memberEnd->asOrderedSet()

-- Verifies if both ends of the association with which a link is typed are navigable
def: isBidirectional : Boolean = associationEnds->select(isNavigable())->size() = 2

-- Verifies if an association with which a link is typed is not navigable
def: isNotNavigable : Boolean = associationEnds->select(isNavigable())->size() = 0

-- Verifies if the association is between two classes
def: isClassClassAssociation : Boolean = associationEnds->select(type.oclIsKindOf(uml::Class))->size()=2

-- Verifies if the association is between two interfaces 
def: isInterfaceInterfaceAssociation : Boolean = associationEnds->select(type.oclIsKindOf(uml::Interface))->size()=2

-- For an association that types a link determines the origine
def: associationStartPoint : Property = 
  if isBidirectional
    then self.associationEnds->at(1) 
  else
    self.associationEnds->select(not isNavigable())->at(1)
  endif

-- For an unidirectional association that types a link determines origine's type
def: associationStartPointType : Type = associationStartPoint.type

-- For an unidirectional association that types a link determines the target
def: associationEndPoint : Property = 
  if isBidirectional
    then self.associationEnds->at(2)
  else
    self.associationEnds->select(isNavigable())->at(1)
  endif
  
-- For an unidirectional association that types a link determines target's type
def: associationEndPointType : Type = associationEndPoint.type

-- Verifies if the association is between a class and an interface and it has this direction
def: isClassInterfaceAssociation : Boolean = associationStartPointType.oclIsKindOf(uml::Class) and associationEndPointType.oclIsKindOf(uml::Interface) 


-- RULE 1 AND RULE 2 

context Connector

-- Definition of link's type 
def: linkType : String = 
  if has2Parts 
    then 'assembly link between parts'
  else 
    if has2Ports 
      then if has1PartWithPort
             then if not port1.isReversed and not port2.isReversed 
                    then 'inbound delegation link between provided ports'
                  else if port1.isReversed and port2.isReversed
                         then 'outbound delegation between required ports'
                       else 'forbidden'
                       endif
                  endif
           else if (port1.isReversed and port2.isReversed) or
                   (not port1.isReversed and not port2.isReversed)
                  then 'forbidden'
                else  'assembly between provided-required ports'           
                endif
           endif
    else  
      if has1PartWithPort
        then if not port1.isReversed
               then 'assembly link between part and provided port'
             else 'assembly link between part and required port'
             endif
      else if not port1.isReversed
             then 'inbound delegation link between part and provided port'
           else 'outbound delegation link between part and required port'
           endif
      endif
    endif        
  endif

-- Rule 1 and Rule 2
inv LinkType: self.linkType <> 'forbidden'


-- HELPER FUNCTIONS

context Connector

-- Verifies if a connector starts from the provided port given as parameter
def: isStartingFromProvidedPort (p:Port) : Boolean = 
  (self.linkType = 'inbound delegation link between provided ports' and p.owner=self.owner) or
   self.linkType = 'inbound delegation link between part and provided port'

-- Verifies if a connector starts from the required port given as parameter
def: isStartingFromReversedPort (p:Port) : Boolean = 
  (self.linkType = 'outbound delegation between required ports' and p.owner<>self.owner) or
   self.linkType = 'assembly between provided-required ports ' or
   self.linkType = 'assembly link between part and required port'

-- Verifies if a connector starts from the port given as parameter
def: isStartingFromPort (p:Port) : Boolean =
  isStartingFromProvidedPort(p) or (isStartingFromReversedPort(p) and p.isReversed)

-- Verifies if a connector starts from a part
def: isStartingFromPart : Boolean = 
  self.linkType = 'assembly link between part and provided port' or 
  self.linkType = 'outbound delegation link between part and required port' or
  self.linkType = 'assembly link between parts'

-- For a port-port connector determines the port from which the link starts
-- For a port-part connector the starting port is the only port of the link 
def: linkStartPort : Port = 
  if has2Ports
    then 
      if isStartingFromPort(port1) 
        then port1
      else port2
      endif
  else port1
  endif

-- For a port-port connector determines the port in which the link ends 
-- For a port-part connector the ending port is the only port of the link
def: linkEndPort : Port =
  if has2Ports
    then if isStartingFromPort(port1) 
           then port2
         else port1
         endif
  else port1
  endif

-- For a port-part connector the starting part is the only part of the link
-- For a part-part connector determines the part from which the link starts (this part it is not navigable)
def: linkStartPart : Property =
  if has1PartAnd1Port
    then part1
  else
    if isEnd1Navigable
      then part2
    else part1
    endif
  endif

-- For a port-part connector the ending part is the only part of the link
-- For a part-part connector determines the part in which the link ends (this part it is navigable)
def: linkEndPart : Property =
  if has1PartAnd1Port
    then part1
  else
    if isEnd1Navigable
      then part1
    else part2
    endif
  endif 

  
-- PROVIDED / REALIZED INTERFACES

context Port

-- Definition of interfaces realized by a port
def: interfaces : Set(Classifier) = self.provided->reject(isInterfaceGroup)

context Classifier

-- Definition of classifier's parents recursive computation
def: getParentsRec : Set(Classifier) =  self.general->union(self.general->iterate(p:Classifier; res:Set(Classifier)=Set{}| res->union(p.getParentsRec)))

context Class

-- Definition of all interfaces directly realized by a class
def: iRealizations : Set(Classifier) = self.interfaceRealization.contract->asSet()

-- Definition of interfaces provided by a class directly or indirectly realized (used in the case of a link not typed by an association)
def: interfaces : Set(Classifier) = 
  iRealizations->union(iRealizations->iterate(i:Interface; res:Set(Classifier)=Set{}| res->union(i.getParentsRec)))
  ->union(self.getParentsRec->iterate(c:Class; res:Set(Classifier)=Set{}| res->union(c.interfaces)))
  ->reject(isInterfaceGroup)

context Interface

-- Definition of interfaces provided by an interface (used in the case of a link typed by an association pointing to an interface)
def: interfaces : Set(Classifier) = self.oclAsType(uml::Classifier)->asSet()->union(self.getParentsRec)->select(not isInterfaceGroup and isInterface)

context Type

-- Determines the set of provided interfaces by a class or an interface
def: interfaces : Set(Classifier) =
  if self.oclIsKindOf(uml::Interface)
    then self.oclAsType(uml::Interface).interfaces
  else
    self.oclAsType(uml::Class).interfaces
  endif


-- RULE 6
  
context Connector

-- Definition of the set of transported interfaces 
def: setTransportedInterfaces : Set(Classifier) =
  if has2Parts
    then Set{OclInvalid}
  else if has2Ports
         then if isTyped
                then if isStartingFromPort(port1)
                       then (port1.interfaces) -> intersection(associationEndPointType.interfaces)
                     else (port2.interfaces) -> intersection(associationEndPointType.interfaces)
                     endif
               else (port1.interfaces) -> intersection(port2.interfaces) 
               endif                              
       else  
         if isTyped
           then (port1.interfaces) -> intersection(associationEndPointType.interfaces)
         else (port1.interfaces) -> intersection(part1.type.interfaces)      
         endif 
       endif
  endif        

-- Rule 6  
inv SetOfTransportedInterfacesNonEmpty: self.setTransportedInterfaces->size()<>0


-- HELPER FUNCTIONS

context Classifier

-- Verifies if the current classifier is compatible with the one given as 
def: isCompatible(c:Classifier) : Boolean =
  if self.oclIsKindOf(uml::Interface) and c.oclIsKindOf(uml::Interface)
    then self.oclAsType(uml::Interface).interfaces->includes(c)
  else
    if self.oclIsKindOf(uml::Class) and c.oclIsKindOf(uml::Interface)
      then self.oclAsType(uml::Class).interfaces->includes(c)
    else
     (c->asSet()->union(c.getParentsRec))->includes(self)
    endif
  endif
  
context Type

-- Verifies if the link end's type (the type of a part) is compatible with the association end's type given as parameter 
def: isCompatible(t:Type) : Boolean = self.oclAsType(uml::Classifier).isCompatible(t.oclAsType(uml::Classifier))

context Port

-- Verifies if the current port is compatible with the association end's type given as parameter 
def: isCompatible(t:Type) : Boolean = self.interfaces->includesAll(t.oclAsType(uml::Interface).interfaces)


-- RULE 3 AND RULE 4

context Connector

-- Verifies if a link starting from a port and typed with an association has the same direction with the association and the interface pointed is included in the set of transported interfaces 
def: linkStartingFromPortVerification : Boolean =
  if isNotNavigable or isBidirectional
    then false
  else 
    if isInterfaceInterfaceAssociation
        then if has1PartAnd1Port
              then linkStartPort.isCompatible(associationStartPointType) and linkEndPart.type.isCompatible(associationEndPointType) and 
                   setTransportedInterfaces->includes(associationEndPointType.oclAsType(uml::Classifier))
             else linkStartPort.isCompatible(associationStartPointType) and linkEndPort.isCompatible(associationEndPointType) and 
                  setTransportedInterfaces->includes(associationEndPointType.oclAsType(uml::Classifier))
             endif
    else false
    endif
  endif
  
-- Verifies if a link starting from a port is well-formed 
def: linkStartingFromPort : Boolean =
  if (not isStartingFromPart) and isTyped
    then linkStartingFromPortVerification
  else true
  endif

-- Rule 3 and Rule 4
inv LinkStartingFromPort: self.linkStartingFromPort


-- RULE 3 AND RULE 5

context Connector

-- For a link between two parts verifies if the ends are compatible with the corresponding ends of the accepted association  
def: linkPartPartVerification : Boolean =
  if isNotNavigable 
    then false
  else
    if isBidirectional
      then (linkStartPart.type.isCompatible(associationStartPointType) and linkEndPart.type.isCompatible(associationEndPointType)) or
           (linkStartPart.type.isCompatible(associationEndPointType) and linkEndPart.type.isCompatible(associationStartPointType)) 
    else (linkStartPart.type.isCompatible(associationStartPointType) and linkEndPart.type.isCompatible(associationEndPointType))
    endif
  endif
  
-- For a link between a part and a port verifies if the ends are compatible with the corresponding ends of accepted association 
def: linkPartPortVerification : Boolean =
  if isNotNavigable or isBidirectional
    then false
  else 
    if isClassClassAssociation
      then false
    else
      if isInterfaceInterfaceAssociation
        then (linkStartPart.type.isCompatible(associationStartPointType) and linkEndPort.isCompatible(associationEndPointType))
      else
        if isClassInterfaceAssociation
          then linkStartPart.type.isCompatible(associationStartPointType) and linkEndPort.isCompatible(associationEndPointType)
        else false
        endif
      endif
    endif
  endif
  
-- Verifies if a link starting from part is typed with an association and if it is well-formed 
def: linkStartingFromPart : Boolean =
  if isStartingFromPart
    then if has2Parts
           then isTyped and linkPartPartVerification 
         else isTyped and linkPartPortVerification
         endif
  else true
  endif

-- Rule 3 and Rule 5
inv LinkStartingFromPart: self.linkStartingFromPart 


-- HELPER FUNCTIONS

context Port

-- Determines all the connectors starting or ending in a port
def: connectors : Set(Connector) = self.end.owner.oclAsType(uml::Connector)->asSet()

-- Verifies if a port has connectors starting or ending in it
def: hasConnectors : Boolean = self.connectors->size()>1

-- Determines all the connectors starting or ending in a port that are not typed with an association
def: connectorsNotTyped : Set(Connector) = self.connectors->select(type.oclIsTypeOf(OclVoid))

-- Verifies if a port has connectors not typed with an association that are starting or ending in it
def: hasConnectorsNotTyped : Boolean = self.connectorsNotTyped->size()>1

-- Determines all the connectors starting from a port (if the boolean parameter is true it collects all the connectors; if it is false it collects only the connectors not typed with association)
def: connectorsStartingFromPort(withType:Boolean) : Set(Connector) = 
  if withType then 
	self.connectors->select(c:Connector|c.isStartingFromPort(self))  
  else
	self.connectorsNotTyped->select(c:Connector|c.isStartingFromPort(self))
  endif

-- Verifies if a port is the origine for at least one connector
def: isStartingPort : Boolean = self.connectorsStartingFromPort(true)->size() >= 1 

-- Determines the union of the sets of transported interfaces on each link starting from the port
def: unionSetForTransportedInterfacesOnLinks(withType:Boolean) : Set(Classifier) =
  self.connectorsStartingFromPort(withType)->iterate(c:Connector; s:Set(Classifier)=Set{} | s->union(c.setTransportedInterfaces))

-- Determines the sum of the number of transported interfaces on each link starting from the port 
def: noOfTransportedInterfacesOnLinks(withType:Boolean) : Integer =
  self.connectorsStartingFromPort(withType)->iterate(c:Connector; i:Integer=0 | i + (c.setTransportedInterfaces->size())) 
  

-- RULE 7
  
context Port

-- Definition of pairwise disjoint sets of transported interfaces 
context Port
def: isPairwiseDisjoint : Boolean = 
  if self.connectorsStartingFromPort(false)->size() >= 2
    then  self.unionSetForTransportedInterfacesOnLinks(false)->size() = self.noOfTransportedInterfacesOnLinks(false)
  else true
  endif

-- Rule 7
inv PairwiseDisjoint: self.isPairwiseDisjoint


-- RULE 8

context Port

-- Verifies if the union of sets of transported interfaces is equal to the interfaces provided/required 
def: isComplete : Boolean =
  if self.isStartingPort
    then unionSetForTransportedInterfacesOnLinks(true) = self.interfaces
  else true
  endif  

-- Rule 8
inv Completeness: self.isComplete       


-- HELPER FUNCTIONS

context Class

-- Determines the number of parts of a composite structure
def: noOfComponents : Integer = self.part->size()

-- Verifies if a class is a composite structure
def: isComposite : Boolean = noOfComponents <> 0

-- Determines the number of active parts owned by a composite structure
def: noOfActiveComponents : Integer = self.part.type.oclAsType(uml::Class)->select(isActive=true)->size()

-- Determines the number of passive parts owned by a composite structure
def: noOfPassiveComponents : Integer = self.part.type.oclAsType(uml::Class)->select(isActive=false and not isProtected)->size()

-- Determines the number of protected parts owned by a composite structure
def: noOfProtectedComponents : Integer = self.part.type.oclAsType(uml::Class)->select(isProtected)->size() 


-- RULE 9 AND RULE 10

context Class

-- Definition of a well-formed class 
def: isWellFormed : Boolean = 
  if self.isActive and isComposite 
    then  if noOfActiveComponents + noOfProtectedComponents = noOfComponents or 
             noOfPassiveComponents = noOfComponents
            then true
          else false
          endif
  else
    if (not self.isActive) and (not isProtected) and isComposite
      then if noOfPassiveComponents = noOfComponents
             then true
           else false
           endif
    else OclInvalid
    endif
  endif

-- Rule 9 and Rule 10
inv CompositeStructure: self.isWellFormed <> false


-- RULE 11

context Class

-- Determines the number of observer parts owned by a composite structure
def: noOfObservers : Integer = self.part.type.oclAsType(uml::Class)->select(isObserver)->size()

-- Definition of a well formed observer
def: isObserverWellFormed : Boolean =
  if self.isObserver and isComposite
    then noOfComponents = noOfObservers
  else true
  endif

-- Rule 11
inv CompositeObserver: self.isObserverWellFormed <> false
\end{lstlisting}
\clearpage 

\end{document}